\documentclass[reprint,amsmath,amssymb,aps,prl,floatfix]{revtex4-2}
\usepackage{natbib}
\usepackage{graphicx}
\usepackage{dcolumn}
\usepackage{bm}
\usepackage{hyperref}
\usepackage[export]{adjustbox}
\usepackage[T1]{fontenc}
\usepackage{float}
\usepackage{todonotes}
\usepackage[normalem]{ulem}
\usepackage{color,xcolor}
\hypersetup{colorlinks=true,citecolor=blue,linkcolor=blue,filecolor=blue,urlcolor=blue,pdfstartview=FitH,pdfpagemode=UseNone}

\begin{document}

\preprint{APS/123-QED}

\title{Disorder information from conductance: a quantum inverse problem}

\author{S. Mukim$^{1}$, F. P. Amorim$^{3}$, A. R. Rocha$^{3}$, R. B. Muniz$^{4}$, C. Lewenkopf$^{4}$ and M. S. Ferreira$^{1,2}$}

 \address{$^1$School of Physics, Trinity College Dublin, Dublin 2, Ireland}
 \address{$^2$Centre for Research on Adaptive Nanostructures and Nanodevices (CRANN) \& Advanced Materials and Bioengineering Research (AMBER) Centre, Trinity College Dublin, Dublin 2, Ireland}
 \address{$^3$ Instituto de F\'{\i}sica Te\'orica, S\~ao Paulo State University, 01140-070, S\~ao Paulo, Brazil}
 \address{$^4$ Instituto de F\'{\i}sica, Universidade Federal Fluminense, Brazil}

\date{\today}

\begin{abstract}
It is straightforward to calculate the conductance of a quantum device once all its scattering centers are fully specified. However, to do this in reverse, {\it i.e.}, to find information about the composition of scatterers in a device from its conductance, is an elusive task. This is particularly more challenging in the presence of disorder. Here we propose a procedure in which valuable compositional information can be extracted from the seemingly noisy spectral conductance of a two-terminal disordered quantum device. In particular, we put forward an inversion methodology that can identify the nature and respective concentration of randomly-distributed impurities by analyzing energy-dependent conductance fingerprints. Results are shown for graphene nanoribbons as a case in point using both tight-binding and density functional theory simulations, indicating that this inversion technique is general, robust and can be employed to extract structural and compositional information of disordered mesoscopic devices from standard conductance measurements.        
\end{abstract}
\maketitle

Structures whose dimensions are comparable to or smaller than the electronic mean free path display transport features not associated with the classical ohmic behaviour \cite{datta_2005}. These are quantum features found by solving the Schr\"odinger equation once the system Hamiltonian is known. Indeed, it is straightforward to describe how  current flows in a quantum device by calculating its conductance once all scattering centres are specified. However, to do this in reverse, {\it i.e.}, to find information about scatterers in a quantum device by simply looking at its conductance is rather challenging. This is a kind of Inverse Problem (IP) which consists of obtaining from a set of observations the causal factors that generated them in the first place. IP are intrinsic parts of numerous visualization tools \cite{medical, fwi, tromp2008spectral, sonar} but are not as common in the quantum realm, and even less so in the presence of disorder. 

Disorder makes the description of the impurity potential by inverse scattering methods quite a daunting task. Multiple scattering depends on the scatterers' locations and therefore likely to affect the electronic dynamics in seemingly unpredictable ways. To make matters worse theory shows that quantum interference in chaotic \cite{Lewenkopf1991} and/or diffusive systems \cite{lee1985universal} gives rise to fluctuations whose statistical properties are universal, {\it i.e.} system independent, indicating that standard IP methods are neither practical nor useful in such situations.

Here we give a different twist to quantum IP approaches and demonstrate that, instead of detrimental, disorder may be actually beneficial to extracting information about scattering centres in a quantum device. In particular, we focus on the energy-dependent conductance of a quantum system, hereafter referred to as the conductance spectrum, which will serve as the only input of the inversion procedure described here.
This is a quantity normally obtained by standard experimental setups of a gated two-terminal device but may also be found by calculation once the underlying Hamiltonian is fully specified.
Here we introduce our inversion methodology by using the latter as a proxy for the former, {\it i.e}, calculated conductance spectra representing their experimental equivalent. The advantage of using calculated input functions is that we can refer back to the Hamiltonian that generated them in the first place, making it possible to assess the success of the inversion procedure.     

\begin{figure}
\centering
\includegraphics[width=\columnwidth]{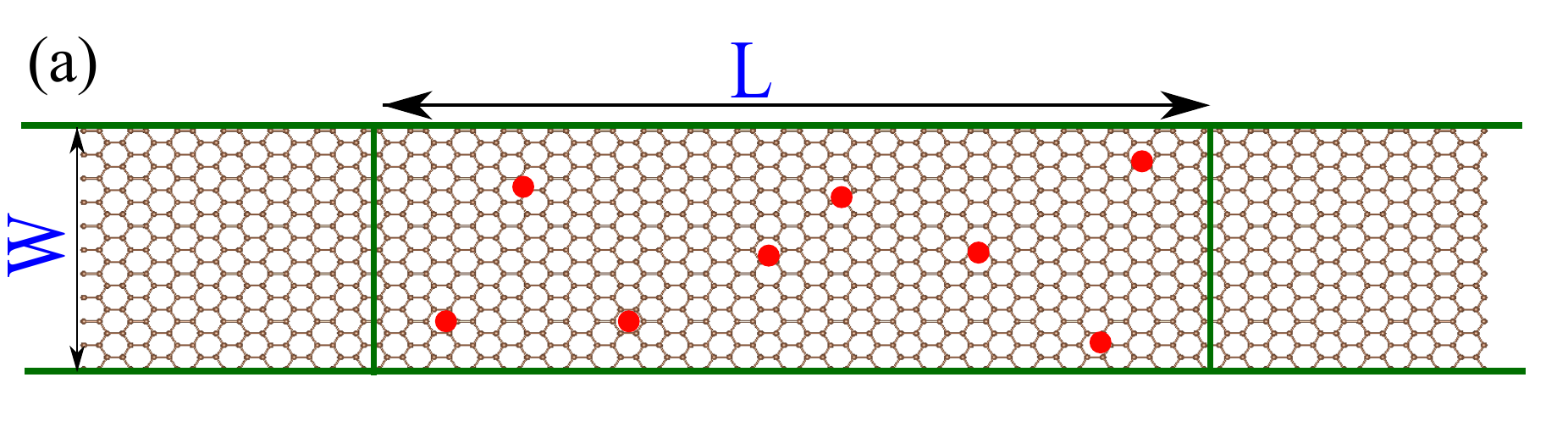}
\includegraphics[width=\columnwidth]{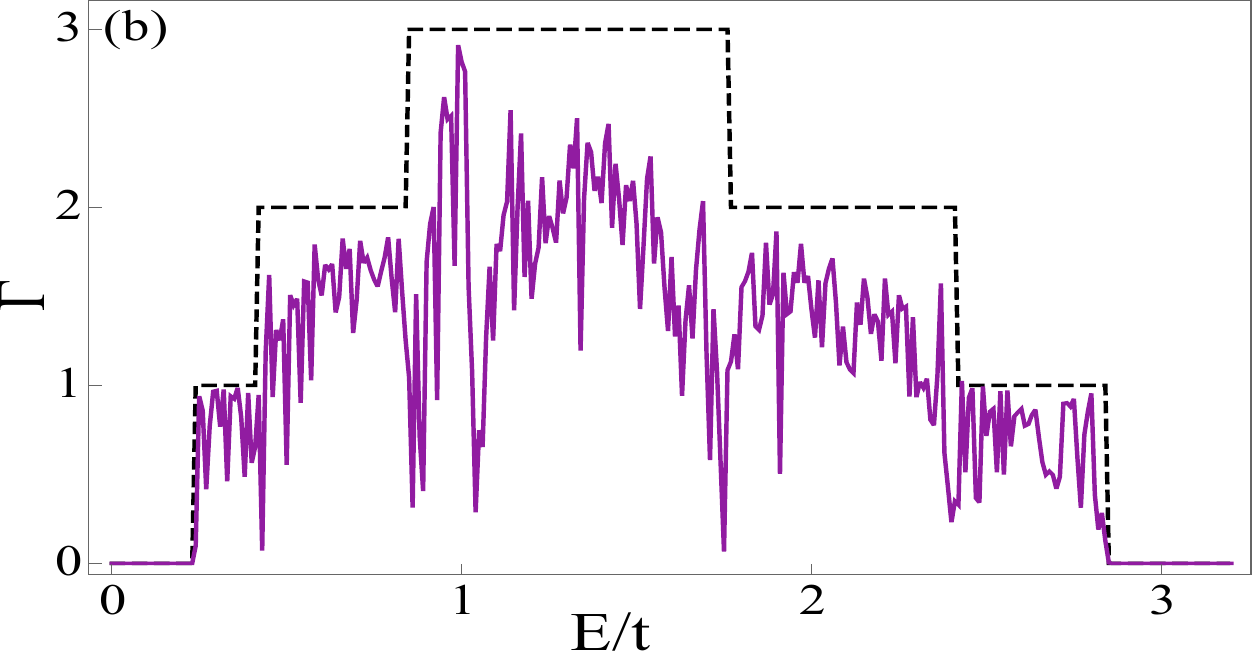}  
\caption{(a) Schematic diagram of the system under consideration. Two semi-infinite leads of width $W$ are separated by a region of length $L$ containing a concentration $n$ of impurities represented by red dots. (b) Calculated conductance for a system made of a graphene nanoribbon of length $L = 100$ unit cells as a function of the chemical potential (in units of $t$). The dashed line corresponds to the conductance of the pristine system, {\it i.e.}, in the absence of impurities. The solid line shows the conductance for a disordered configuration containing $N=42$ substitutional impurities ($\epsilon=0.5t$), which accounts for a concentration of $n=3\%$.}
\label{fig1}
\end{figure}

Let us start by defining the system to be used throughout the manuscript. It consists of 
two electrodes separated by a scattering region of length $L$ and width $W$, as illustrated in Fig.~\ref{fig1}(a), a rather typical setup of electronic transport. The distinction between the leads and the central region is that the latter contains impurities as scattering centres \cite{tapaszto2008tailoring}. 
In the linear response regime, the Landauer conductance reads
${\cal G} = 2 e^2/h \int dE (-\partial f/\partial E) \Gamma (E)$, where $f(E)$ is the Fermi distribution and $\Gamma(E)$ is the dimensionless conductance (or transmission) given by \cite{meir1992landauer,datta_1995}
\begin{equation}
    \Gamma(E) = {\rm tr} \big[ {\bf G}^r(E) {\bm \Gamma}_R(E) {\bf G}^a(E) {\bm \Gamma}_L(E)\big]\,
    \label{eq-gamma}
\end{equation}
Here ${\bf G}^r$ (${\bf G}^a = [{\bf G}^r]^\dagger)$ is the full retarded (advanced) Green function and ${\bm \Gamma}_L$ (${\bm \Gamma}_R$) is the line width function accounting for the injection and lifetime of states in the left (right) contact. For simplicity, we consider the zero-temperature limit ${\cal G} = 2e^2 /h \, \Gamma(E)$. Thermal corrections will be shown to have little effect on our procedure.   

The expression for $\Gamma$ in Eq.~(\ref{eq-gamma}) is model-independent, {\it i.e.}, once the Hamiltonian is known one can find the corresponding Green function and obtain the energy dependent conductance of the system \cite{meir1992landauer}.
We focus on systems with relatively simple electronic structures, namely graphene nanoribbons (GNR). Although not an essential requirement, it helps to illustrate the methodology since GNR are well described by the tight-binding model \cite{neto2009electronic,DasSarma2011}. In this case the nearest-neighbour hopping $t$ and the on-site energies fully define the electronic structure of the nanoribbon. 
Fig.~\ref{fig1}(b) shows the energy-dependent conductance for an armchair-edged GNR of width $W = 3a$, $a= 2.7$\AA\rm{\,} 
being the graphene lattice parameter. The dashed line is the conductance spectrum $\Gamma_0$ for the pristine GNR. Results are shown for positive energies knowing that $\Gamma_0(E) = \Gamma_0(-E)$. 

We now introduce $N$ substitutional impurities. It is convenient to express the impurity number also as the percentage concentration $n$ defined as $n=100 \times (N/N_{\rm tot})$, where $N_{\rm tot}$ is the total number of sites in the central region. Both $N$ and $n$ will be used interchangeably. The scattering strength of the
impurities is characterized by the contrast between their on-site potential $\epsilon$ relative to that of the host, chosen to be zero. The solid line of Fig.~\ref{fig1}(b)
shows the conductance for the GNR with $N = 42$ impurities, $\epsilon=0.5\,t$ and $L=100$ unit cells ($n=3\%$). Impurity locations were randomly selected but kept fixed in the underlying Hamiltonian that generated the conductance $\Gamma(E)$ of Fig.~\ref{fig1}(b).
 
Being the conductance sensitive to the locations of scattering centres, it is difficult to devise an inversion tool capable of spatially mapping all impurities from the conductance spectrum information alone. The brute-force method of comparing the input conductance
$\Gamma(E)$ with those of every possible disorder configuration is not viable because the number of combinations is too large for any practical situation. Machine-learning strategies are currently being attempted to overcome this combinatorial hurdle \cite{lopez2014modeling,hansen2013assessment,mach-learn0,mach-learn1,mach-learn2,mach-learn3,mach-learn4,mach-learn5} but spatial mapping of quantum devices through inversion remains challenging.   

Nevertheless, given the conductance spectrum $\Gamma(E)$ of a device, one might ask whether it is possible to find the exact number of scattering centres in it. This may not reveal the position of every impurity but it is a valuable piece of information and a lot more feasible to obtain. While the brute-force approach remains impractical, we must account for as many disorder realizations as possible. We define the configurationally averaged (CA) conductance $\langle \Gamma \rangle$ by summing over $M$ realizations with the same number $N$ of impurities (or concentration $n$), {\it i.e.},
\begin{equation}
     \langle \Gamma(E,n)\rangle = \frac{1}{M}\sum_{j=1}^M \Gamma_j(E)\,\,,
     \label{config-avg}
\end{equation}
where $j$ labels the different configurations. See SM \cite{SM} for a discussion on the suitable choices for $M$. 

\begin{figure}
    \includegraphics[width=0.85\columnwidth]{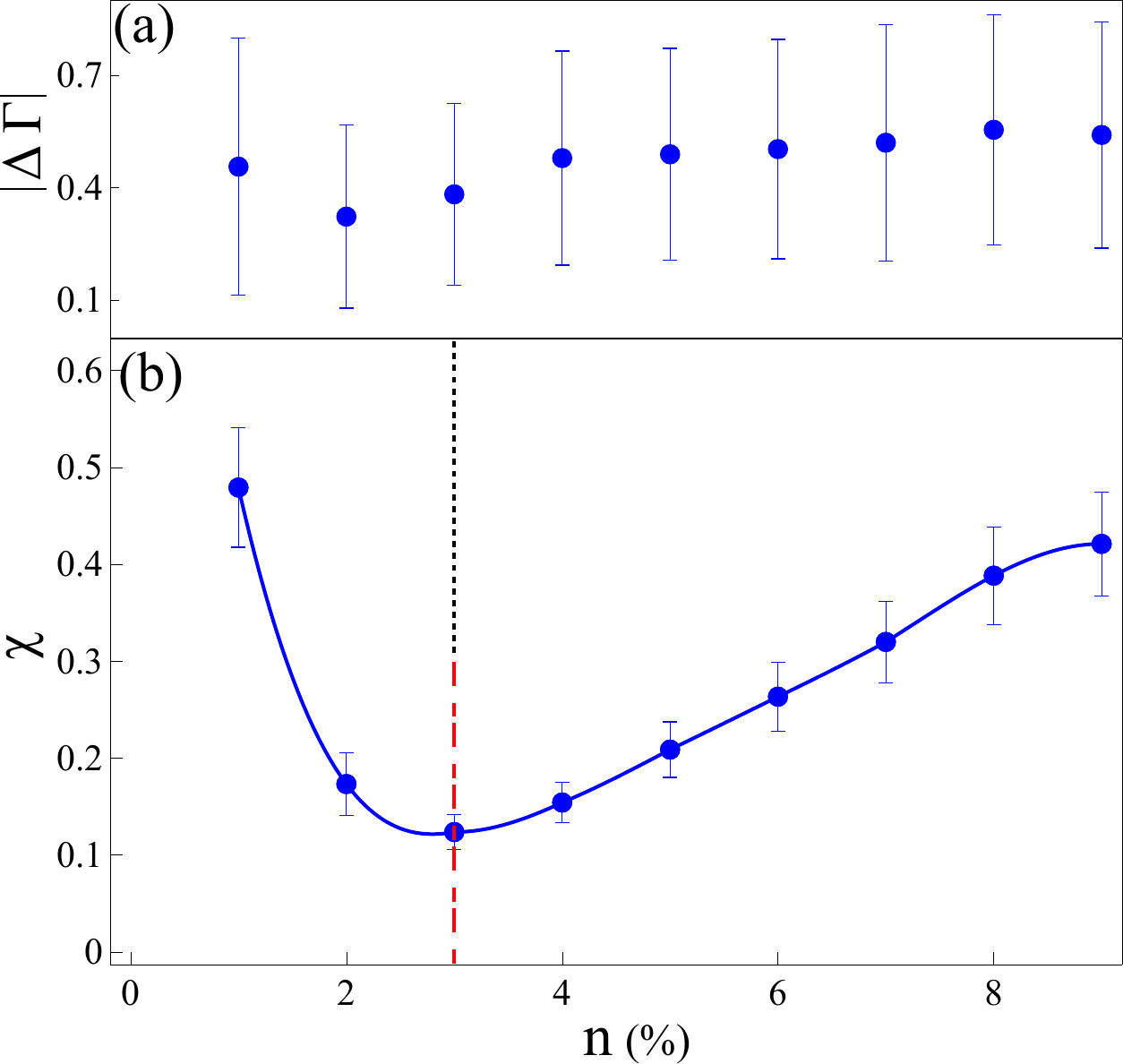}
    \caption{(a) Absolute value of the conductance deviation $\vert\Delta \Gamma(E,n)\vert$ as a function of the impurity concentration $n$ (in percentage) for a fixed energy ($E=0.42 t$). (b) Misfit function $\chi$ (in arbitrary units) as a function of $n$. The vertical (red) dashed line on the lower part of the panel indicates the real number of impurities used to generate the sample conductance, which coincides with the minimum of $\chi(n)$. The (black) dotted line in the upper part of the panel is the approximate concentration $n^*$. Integration limits were ${\cal E}_-=0.5 t$ and ${\cal E}_+ = 1.5 t$ and the solid (blue) line is simply a guide to the eyes.}
    \label{plot:misfit}
\end{figure}

The deviation between an arbitrary conductance result and its CA counterpart is given by 
\begin{equation}
    \Delta \Gamma(E,n) = \Gamma(E) - \langle \Gamma(E,n)\rangle
    \label{dev}
\end{equation}
We reiterate that $\Gamma(E)$ acts as the input conductance spectrum of a single realization and represents the immutable conductance of the device under investigation. The CA conductance spectrum, on the other hand, reflects the contribution of very many configurations and depends, in addition, on the impurity concentration. By treating $n$ as a variable parameter, we can look for minimization trends in $\Delta \Gamma(E,n)$ that might indicate the real concentration in the device. Unfortunately, when plotted as a function of $n$ in Fig.~\ref{plot:misfit}(a), the deviation $\Delta \Gamma$ for a fixed energy $E$ is featureless with wide error bars that result from repeating the calculation 1000 times. 

However, much cleaner trends are seen when $\Delta \Gamma$ is used in the form of a functional that measures how good a match $\Gamma(E)$ and $\langle \Gamma(E,n) \rangle$ are. This quantity is the misfit function $\chi(n)$ defined as 
\begin{equation}
    \chi(n) =  \int_{{\cal E}_-}^{{\cal E}_+} dE \, \left[\, \Gamma(E) - \langle \Gamma(E,n)\rangle \, \right]^2 \,\,,
    \label{eq:misfit}
\end{equation}
where ${\cal E}_-$ and ${\cal E}_+$ establish the energy window over which the integration takes place. Fig.~\ref{plot:misfit}(b) shows $\chi$ as a function of $n$ and displays a more distinctive trend with smaller error bars. The plot indicates that there is a sweet spot in impurity concentration for which the integrated deviation is minimal. Remarkably, this agrees with the actual number of impurities used in the calculation of $\Gamma(E)$, shown as a vertical (red) dashed line in the lower part of Fig.~\ref{plot:misfit}(b). Such a coincidence suggests that it might be possible to use $\chi(n)$ as an inversion tool to find the number of impurities in a quantum device from simple conductance measurements. Note that no prior knowledge of the actual number of impurities was necessary to identify the misfit-function minimum. 

To demonstrate that the agreement between the minimizing concentration of $\chi$ and the actual impurity concentration is not a fortuitous coincidence, we proceed to write the CA conductance \cite{datta_1995} to linear order in $n$ 
\begin{equation}
\langle \Gamma(E,n) \rangle = \Gamma_0(E) - \beta(E) \, n,
\label{CA-approx}
\end{equation}
where $\beta(E)$ is the derivative of $\langle \Gamma(E,n) \rangle$ with respect to $n$ 
evaluated at $n=0$. The misfit function $\chi(n)$ will naturally develop a minimum at $n^*=-B/A$, where 
\begin{equation}
A = \int_{{\cal E}_-}^{{\cal E}_+} \!\!dE  \beta^2(E)
\;\ \mbox{and} \;
B = \int_{{\cal E}_-}^{{\cal E}_+} \!\!dE   \beta(E)  \big[\Gamma(E) - \Gamma_0(E)\big] .
\label{B}
\end{equation}
The vertical (black) dotted line shown on the upper part of Fig.~\ref{plot:misfit}(b) indicates the value of $n^*$. Note that both the dotted and dashed lines are aligned, {\it i.e.}, $n^*$ coincides not only with the actual concentration $n$ but also with the impurity concentration that minimizes $\chi(n)$. In fact, $n^*$ is an excellent approximation up to 3\% impurity concentration and provides a simple yet accurate way of identifying the misfit function minimum. \footnote{An alternative definition of $\beta(E)$, shown in the SM \cite{SM}, may extend the validity of the approximation to up to 7\%.}

\begin{figure}
    \centering
    \includegraphics[width=\columnwidth]{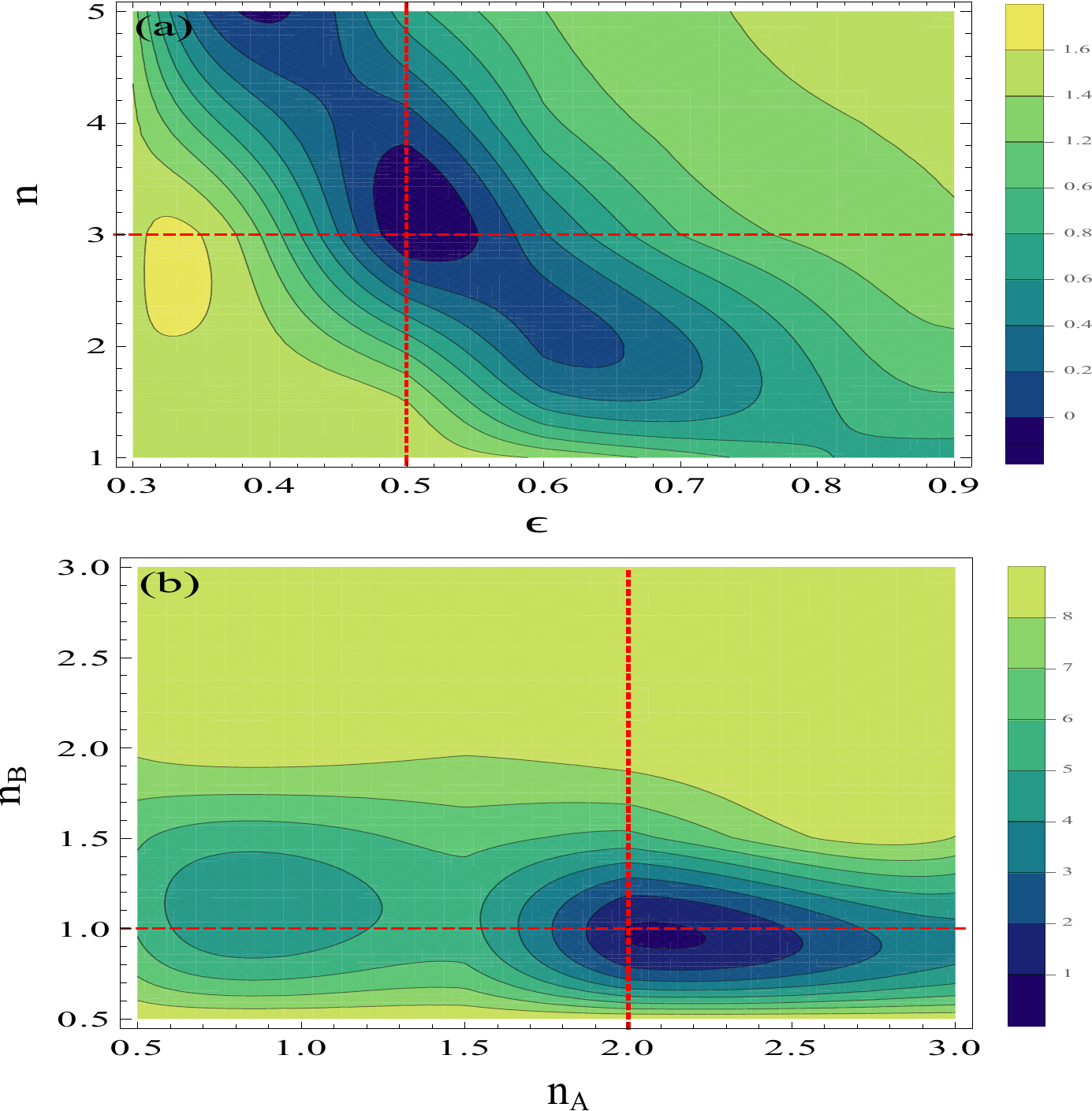}
    \caption{2D contour plots of the logarithm of the misfit function. Dashed lines in the plots indicate the values of the respective quantities used in the underlying Hamiltonians. (a): $\chi$ as a function of $n$ and $\epsilon$. (b): $\chi$ as a function of $n_a$ and $n_b$.}
    \label{plot:chi-n/eps}
\end{figure}

Besides the impurity concentration, other degrees of freedom can be added to the IP in question. Let us now consider an arbitrary sample conductance but this time assume that nothing is known about the impurity, {\it i.e.}, neither its concentration $n$ nor its on-site potential $\epsilon$. In this case, the input $\Gamma(E)$ is simply treated as the conductance of a system with an unknown number of unspecified impurities. To calculate the misfit function we must compute the CA conductance which will now depend on $\epsilon$ as well as $n$. A 2D contour plot of $\chi$ as a function of these two quantities is shown in Fig.~\ref{plot:chi-n/eps}(a).
A distinctive minimum is seen which, once again, coincides with the exact values used to generate the input conductance, shown as dashed lines.
Consequently, both the type and concentration of scattering centres inside a quantum device can be identified through its energy-dependent conductance fingerprints. Furthermore, still using two degrees of freedom, the IP can also be implemented in the case of two types of impurities with unknown concentrations. In other words, two impurities described by known on-site potentials $\epsilon_A$ and $\epsilon_B$ are randomly dispersed with respective concentrations $n_A$ and $n_B$, which are unknown. Writing the CA conductance as a function of both concentrations leads to the misfit function being plotted now as a function of the same quantities in Fig.~\ref{plot:chi-n/eps}(b). The dashed lines stand for the actual values $n_A$ and $n_B$. They accurately match the concentration values that minimize the misfit function. This indicates that it is possible to increase the number of degrees of freedom in the inversion procedure, offering variety and versatility in how we wish to interrogate the system. While such an increase may lead to the appearance of more minima in the misfit function, it remains a straightforward numerical task to identify them all.   
\begin{figure}
    \includegraphics[width=0.85\columnwidth,height=1\columnwidth]{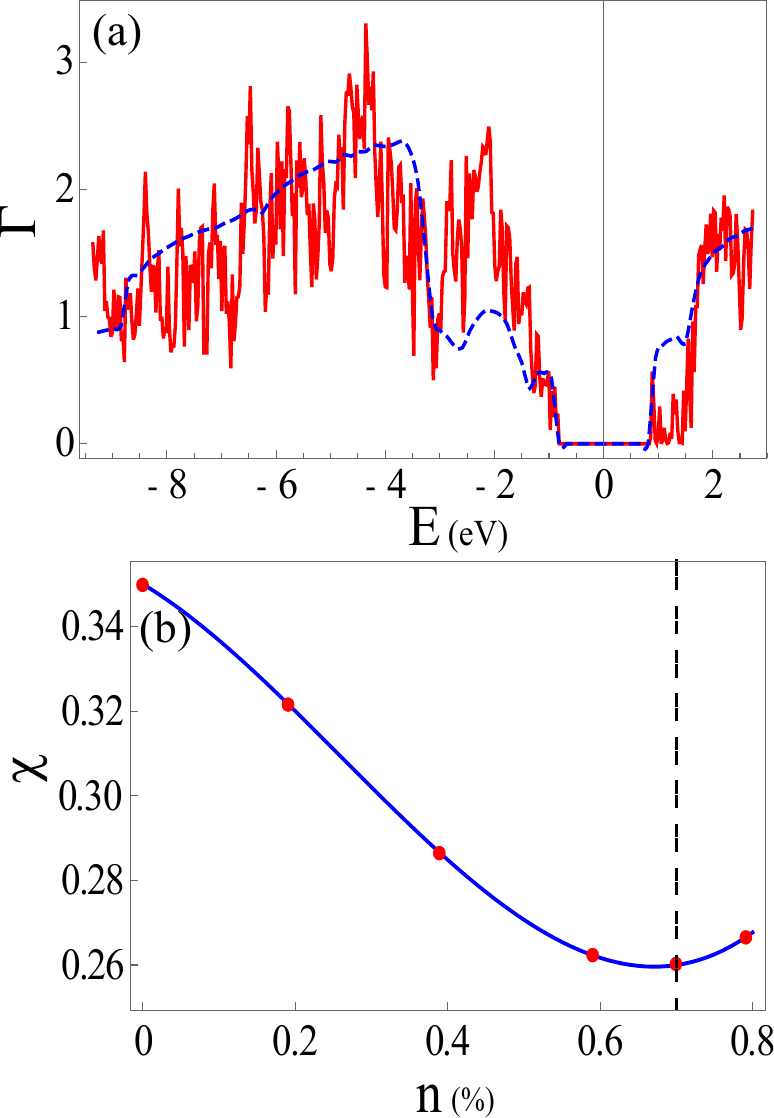}
\caption{(a) The (red) solid line is the conductance spectrum $\Gamma(E)$ obtained from a specific disorder realization of $N=10$ nitrogen impurities, {\it i.e.}, $n=0.7\%$. The (blue) dashed line represents the CA conductance ($M=2000$) calculated for the concentration that minimizes the misfit function $\chi$, shown in (b) as a function of $n$. The misfit function was generated with an energy window defined by ${\cal E}_-=0$ and ${\cal E}_+=2 \, {\rm eV}$. Dashed line indicates the exact concentration used to generate the input conductance.}
    \label{plot:dft}
\end{figure}

To show that this method is indeed model-independent we have also tested it in the case of a system fully described by Density Functional Theory (DFT). In particular, DFT-based conductance calculations were performed for armchair GNR of sizes $L = 100$ unit cells and $W=3a$ containing a specific spatial distribution of $N=10$ nitrogen atoms as substitutional impurities, which corresponds to $n=0.7\%$. See SM \cite{SM} for details of the DFT calculations. The solid line in Fig.~\ref{plot:dft}(a) shows the conductance spectrum $\Gamma(E)$ of the system obtained within DFT, which will now serve as the input conductance for the inversion procedure. Regarding the CA conductance, there are two options on how $\langle \Gamma(E,n)\rangle$ can be obtained. First option is to repeat the steps taken to generate $\Gamma(E)$ of Fig.~\ref{plot:dft}(a) over several different disorder configurations. Bearing in mind that the impurity concentration $n$ must be kept as a variable parameter, the averaging procedure defined in Eq.~(\ref{config-avg}) requires $M$ different configurations for every single value of $n$, which indicates how computationally demanding this task might become if carried out entirely within DFT. The alternative option is to use the tight-binding (TB) model to carry out the CA calculation. In this case the TB model provides a fast averaging strategy without necessarily compromising in accuracy.

Here, we have selected the latter option and made use of DFT conductance spectra to identify suitable TB parameters that describe the nitrogen impurities in the GNR. See SM \cite{SM} for details, including a brief discussion on the choices available to extract TB parameters from DFT calculations \cite{gresch2017z2pack, Pizzi2020, nardelli2018paoflow,vanderbilt_2018} and their implementation in large scale transport calculations \cite{groth2014kwant,Lima2018,joao2020kite}. The CA conductance $\langle\Gamma(E,n)\rangle$ is then generated within the TB model employing $M=2000$ realizations, leading subsequently to the misfit function $\chi(n)$ of Fig.~\ref{plot:dft}(b). Once again, $\chi(n)$ displays a minimum at exactly the same nitrogen concentration used to generated the input conductance $\Gamma(E)$. Another indication of success can be seen by plotting the CA conductance evaluated at the minimizing concentration of $\chi(n)$. Shown as a dashed line in Fig.~\ref{plot:dft}(a), it has all the key features of $\Gamma(E)$ even though both curves were calculated independently.  Despite the computational complexity of obtaining the misfit function entirely within DFT, this has also been tested and, reassuringly, we find exactly the same answer (see SM \cite{SM} for details). 

All input spectra in this manuscript were calculated from known Hamiltonians because it is then straightforward to test the success of the inversion method. Ultimately, inversions must be performed based on experimental conductance data of systems for which we do not have the full Hamiltonian. This calls for an inversion tool that is reliable, general and robust. The fact that our inversion strategy works for systems whose electronic structures are described by a simple TB model as well as by DFT calculations is indicative of the generality and robustness of this approach. 

That disorder is beneficial to this inversion procedure is made evident by the distinction between the two panels of Fig.~\ref{plot:misfit}. Viewed at a fixed energy, the deviation between the input and the CA conductance spectra reveals very little about the system. In contrast, the misfit function entails a lot of information.
The efficiency of the procedure relies on an ergodic hypothesis: conductance fluctuations of a single sample versus energy are related to sample to sample fluctuations at a fixed energy. More precisely, the ergodic hypothesis assumes that a running average over a continuous parameter upon which the conductance depends is equivalent to sampling different impurity configurations. Here the continuous parameter is the energy, but the concept can be extended to a range of other quantities, {\it e.g.}, magnetic field, gate voltage, etc. A thorough mathematical discussion of this issue is beyond the scope of this manuscript, but it is worth mentioning that the ergodic hypothesis can be proven exactly for certain models of disordered systems\cite{French1978,Brezin1994,Guhr1998}. That explains why the energy integration induces a distinctive minimum in $\chi(n)$ since it is analogous to considering a wider universe of disorder configurations in the CA procedure. The success of the inversion procedure will therefore depend on how wide the integration range is. See SM \cite{SM} for details on the suitable choices for integration limits.

Another feature of the conductance fluctuations of quantum devices \cite{lee1985universal,lee1987universal} is that the correlation function $C(\delta X)= \langle \Gamma(X+\delta X/2) \Gamma(X - \delta X/2)\rangle - \langle \Gamma(X)\rangle^2$ is also universal \cite{Brezin1994}. The latter is characterized by a correlation length $\delta X_{\rm cor}$ that is system dependent. 
Therefore, the method works best when the integration interval $\Delta X$ is such that $\Delta X/\delta X_{\rm cor} \gg 1$.
In our case, where we consider variations of the system chemical potential, $\delta E_{\rm cor}$ scales with the mean level density times the transmission 
\cite{Alves2002}, which makes $\delta E_{\rm cor}$ much larger than typical experimental temperatures, justifying our earlier claim that temperature has little effect on our inversion procedure. This also explains why the energy window used in the misfit-function definition does not have to be very large.   

In summary, the inversion procedure presented here provides a mechanism capable of identifying the composition of scattering centres in a quantum device by simply looking at the energy dependence of the two-terminal device conductance. Assuming that the impurity type within the device is known, the procedure establishes the exact impurity concentration in it. Alternatively, if no information is known {\it a priori} about the scatterers, the inversion identifies their scattering strength and respective concentration. Finally, with a mixture of two different types of impurities we are able to establish the fractional concentration of each component of the device. Despite being presented with GNR, the technique is not material-specific and performs remarkably well in the ballistic, diffusive and at the onset of localised transport regimes. The method is based on the notion that conductance fluctuations carry little system-specific information, the average conductance depends smoothly on the variables of interest and on a standard ergodic hypothesis \cite{French1978}, requiring only an effective (single-particle) description of the system. 
\footnote{ The applicability of the method for (strongly) interacting systems, {\it i.e.} beyond the mean field approximation, is not clear yet.} This small set of conditions renders a very robust and versatile methodology that can extract structural and compositional information of quantum devices from standard transport measurements. 

\acknowledgments
The authors acknowledge support from \-FAPESP\- (Grant \# \-FAPESP\-  2017/02317-2, 2016/01343-7, and 2017/10292-0), CNPq (Grants 308801/2015-6 and 312716/2018-4), FAPERJ (Grant E-26/202.882/2018), and the ICTP-Simons Foundation Associate Scheme. The authors also thank the National Laboratory for Scientific Computing (LNCC/MCTI, Brazil) for providing HPC resources of the SDumont supercomputer, which have contributed to the results reported here. Discussions with S. R. Power and M. Kucukbas are greatly acknowledged. 

\bibliography{mybib,dftbib}

\end{document}


\preprint{APS/123-QED}

\title{Disorder information from conductance: a quantum inverse problem \\ (Supplemental Material)}

\author{S. Mukim, F. P. Amorim, A. R. Rocha, R. B. Muniz, C. Lewenkopf, and M. S. Ferreira}
\maketitle

\section{I. Inversion procedure: Technical aspects}

Here we provide details of the inversion procedure implementation and an analysis of its accuracy. For that purpose we use the model system introduced in the main text as an illustrative example. Our strategy relies on calculating the conductance spectra that is used as input functions of the inversion method simply because we can refer back to the Hamiltonian that generated them in the first place, allowing us to establish whether or not the inversion was successful. A simple test is to start from the Hamiltonian of an arbitrary specific configuration of $N$ impurities randomly located and calculate the corresponding conductance spectrum. This translates into an impurity concentration $n$ (see main manuscript). The inversion works by finding the concentration that minimizes the misfit function, which we call $n_{min}$. We then define the inversion procedure accuracy as $\alpha=\vert n-n_{min}\vert/n$. In what follows we analyse how $\alpha$ depends on a few essential parameters and show that a careful tuning of such parameters may improve the success rate of the inversion methodology. 

\subsection{1. Accuracy of the inversion procedure}

\begin{figure}
    \centering
    \includegraphics[width=\columnwidth]{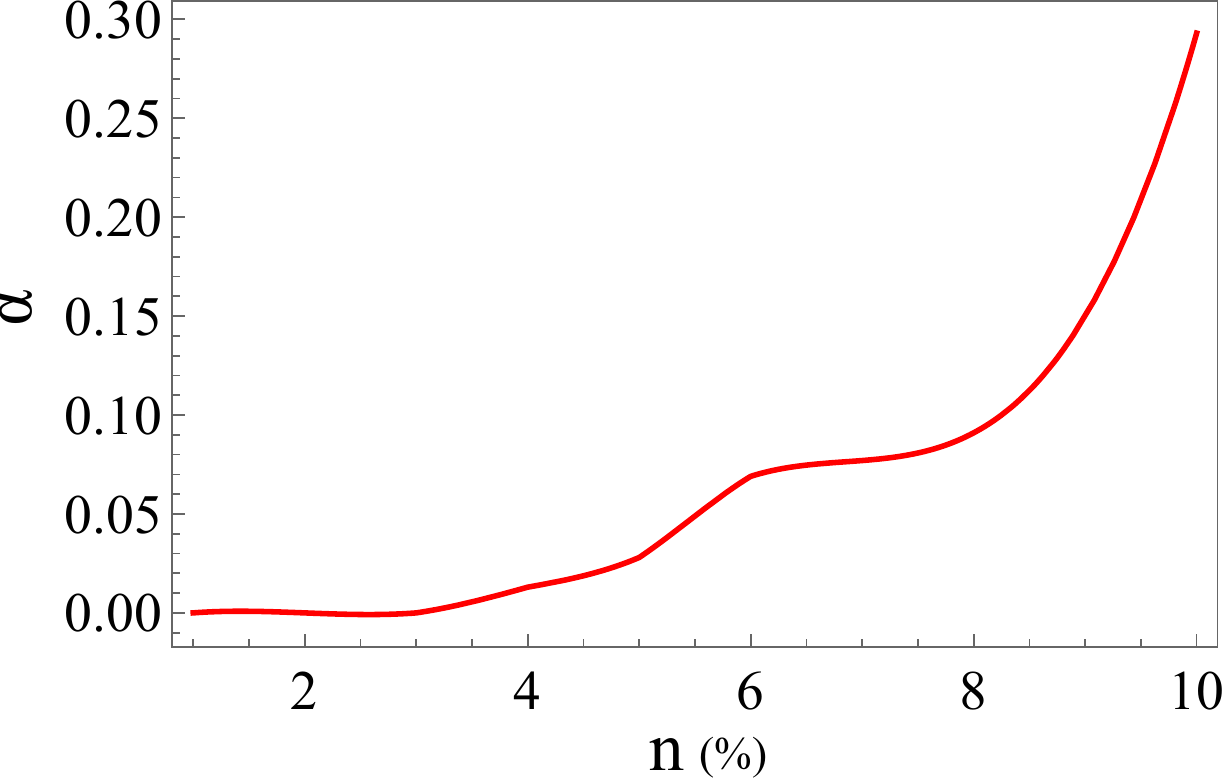}
    \caption{Inversion accuracy $\alpha$ plotted as a function of the impurity concentration $n$ for a GNR with $W=3a$ and $L=100$ unit cells. The impurity on-site potential was arbitrarily chosen as $\epsilon=0.5 t$.}
    \label{error_n}
\end{figure}

Figure~\ref{error_n} shows the inversion accuracy $\alpha$ for graphene nanoribbons (GNR) with $W=3a$ and $L=100$ unit cells as a function of impurity concentration. One can see that $\alpha$ is very small for all concentrations below 8\%, clearly indicating that our inversion tool is very reliable for dilute impurity concentrations.

It is worth recalling that we can not only confirm what the real impurity concentration that generated the Hamiltonian is but also the exact locations of the randomly placed impurities. Interestingly, as the concentration increases we find a growing number of impurities that are adjacent to one another forming pairs that may act as if they were isolated scatterers. For example, a randomly selected configuration of $15$ impurities may have $13$ of them well isolated whereas the remainder are clustered into a pair. In cases like that the inversion finds a concentration that would correspond to $14$ impurities for the simple fact that the cluster may actually behave as a single scatterer. This explains why the inversion error also increases with the concentration. It is possible to reduce the error further by accounting for the probability of forming such pairs. By relaxing the assumption of short-range uncorrelated disorder originally made, we may adopt a similar strategy to the one used to produce Figure 3(b) of the main text, {\it i.e.}, where we treat impurity pairs as if they corresponded to a different class of impurities. In this way it is straightforward to discern between isolated and paired impurities, leading to an improved accuracy that can be sustained up to higher concentrations. 

\subsection{2. Averaging energy window}

The misfit function $\chi$ depends on the  integration limits ${\cal E}_+$ and ${\cal E}_-$, that is, the averaging energy window. 

\begin{figure}
    \centering
    \includegraphics[width=\columnwidth]{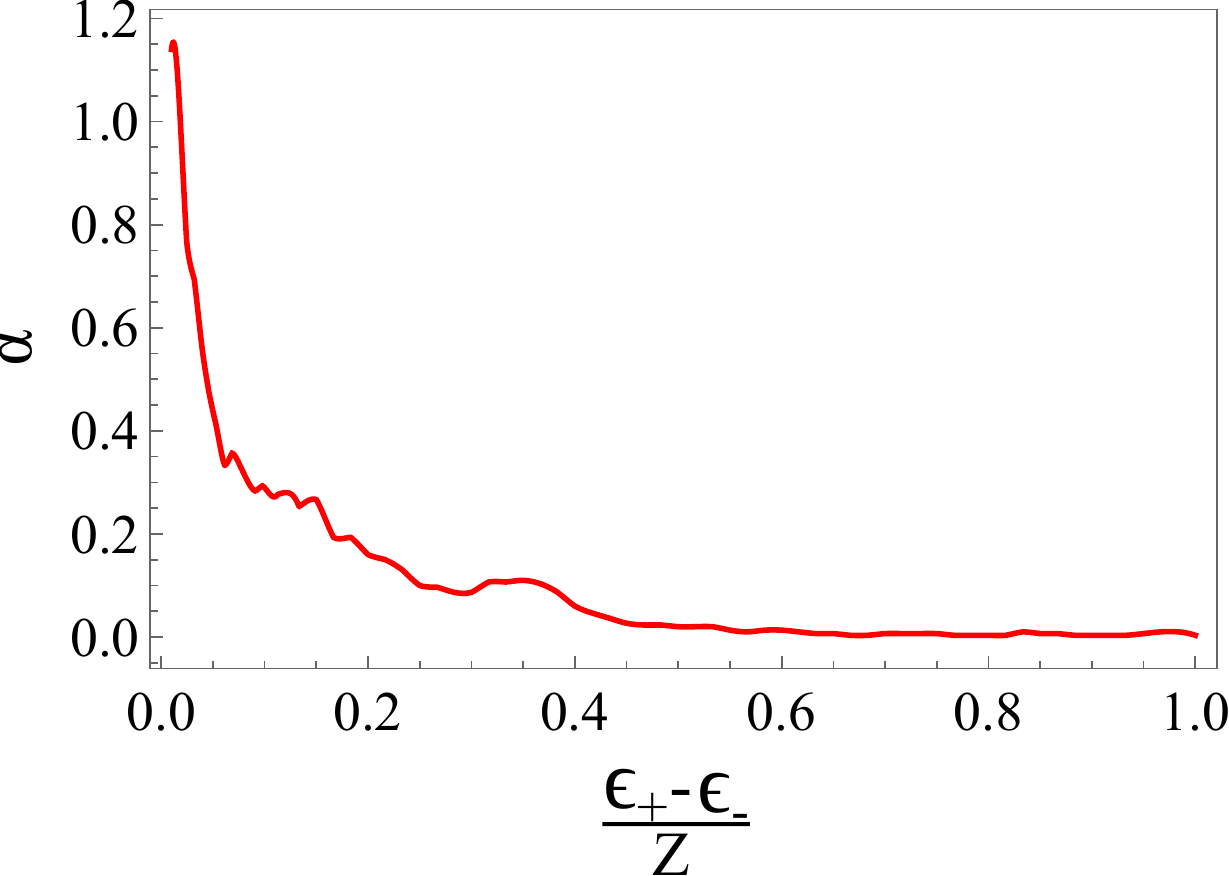}
    \caption{Inversion accuracy $\alpha$ as a function of averaging energy window considered for inversion technique. Averaging is carried out with 100 different configurations of $L=100$ unit cells, each one of them containing 3\% of impurities. }
    \label{error_bw}
\end{figure}

Figure \ref{error_bw} plots the inversion error $\alpha$ as a function of $({\cal E}_+ - {\cal E}_-)/Z$, where $Z$ is the bandwidth. Note that $\alpha$ is large for very narrow energy windows, in perfect agreement with the single-energy featureless results of Fig.~2(a) presented in the main manuscript. Note that any window at least 20\% of the bandwidth generates excellent results. 

In practical situations, it can be a difficult task to obtain the conductance spectrum over a significant portion of the bandwidth. We argue that good accuracy, {\it i.e.}, small values of $\alpha$, can be obtained with fairly small averaging windows. The idea of averaging is to eliminate the universal conductance fluctuations from the misfit function $\chi$. 
Hence, the averaging window has to be sufficiently broad to contain a large number of oscillations of $\Gamma(E)$. 
More precisely, $\Delta E/\delta E_{\rm cor}\gg 1$, where $\delta E_{\rm cor}$ is the transmission autocorrelation length. 
Fortunately, the conductance fluctuations in mesoscopic systems are characterized by $\delta E_{\rm cor}$ that are typically much smaller than the band width $Z$. Theory \cite{Lewenkopf1991,Alves2002} shows that $\delta E_{\rm cor}$ depends essentially on the mean level density and on the average transmission $\langle\Gamma(E,n)\rangle$. We point out that the same idea can be extended to variations other than the energy, such as with external fields, strain, etc.

\subsection{3. Device size and transport regime} 

\begin{figure}
    \centering
    \includegraphics[width=\columnwidth]{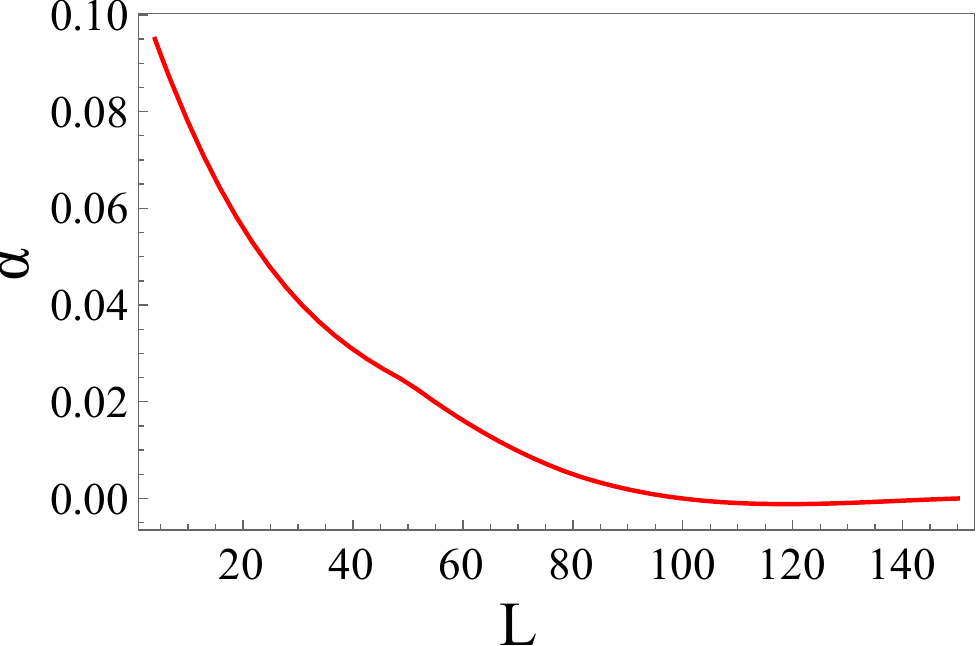}
    \caption{Inversion accuracy $\alpha$ as a function of device size $L$ (in unit cells). Devices of all sizes contained a fixed impurity concentration of $n=4\%$.}
    \label{error_l}
\end{figure}

It is important to assess how accurate the inversion procedure is in different transport regimes. Fig. \ref{error_l} plots the inversion accuracy as a function of the device size $L$ for a concentration of $n=4\%$. It is unquestionable that the accuracy is very good ($\alpha < 0.1$) for all sizes considered. It is instructive to put this result in context by looking at the localization length of the system, which in the case of $n=4\%$ of impurities is approximately $100$ unit cells. That means that the inversion performs remarkably well regardless of the transport regime, i.e., whether it is ballistic, diffusive or at the onset of the localized regime.        

\subsection{4. Averaging procedure and statistical significance of the Configuration Average (CA) calculations}

As discussed in the main text, the underlying hypothesis for the applicability of the method are that  conductance fluctuations contain very little system-specific information (depending on disorder they can even have a universal character) and that the average conductance depends smoothly on the variables of interest. To obtain accurate averages from a data set of the conductance as a function of a continuous (running) variable we use a standard ergodic hypothesis \cite{French1978}. To be more precise, let us define the running average of a quantitity of interest 
$O$ as 
$$
\langle O(X) \rangle \equiv \frac{1}{\Delta X} \int\! dX  \, O(X)
$$
where $X$ is a generic continuous (running) variable and $\Delta X$ is its corresponding range in the data set. In our study we employ the ergodic hypothesis to replace averages over configurational disorder (ensemble averages) with running averages, {\it i.e.},  
$$
\overline{ \Gamma(X) } = \langle  \Gamma(X) \rangle
$$
and
$$
\overline{ \Gamma(X) \Gamma(X + \delta X)} = 
\langle \Gamma(X) \Gamma(X + \delta X) \rangle 
$$
Note that the conductance is a two-point Green's function. This allows us to justify our statistical assumptions based on the vast literature on both the Kubo conductity for disordered diffusive bulk systems and Landauer conductance for disordered/chaotic nanostructures.

Regarding the number of configurations needed in the CA procedure, as standard, the higher the value of disorder realizations $M$ the smaller the fluctuations in $\langle \Gamma(E,n) \rangle$. 
Universal conductance fluctuations (UCF) help one to obtain accurate results for  $\langle \Gamma(E,n) \rangle$ at a modest value of $M$. Both for diffusive \cite{lee1985universal,lee1987universal} and chaotic ballistic systems \cite{Baranger1994}, var($\Gamma) \approx 1$ is the main fingerprint of UCF. 
Hence, the CA relative statistical error is expected to scale with $[\sqrt{M}\times \langle \Gamma(E,n) \rangle]^{-1}$.

To bring the degree of fluctuations to an acceptable level and produce results that are statistically significant, we note that $M$ is usually of the order $10^3$. \cite{Alhassid2000,lopez2014modeling}. 
One-dimensional (1D) systems, with corresponding smaller mean dimensionless conductance, require higher values of $M$ to achieve statistical significance. Likewise, larger 2D systems and 3D ones, with larger mean transmission coefficients than those considered in this study, may require fewer configurations.

\begin{figure}
    \centering
    \includegraphics[width=\columnwidth]{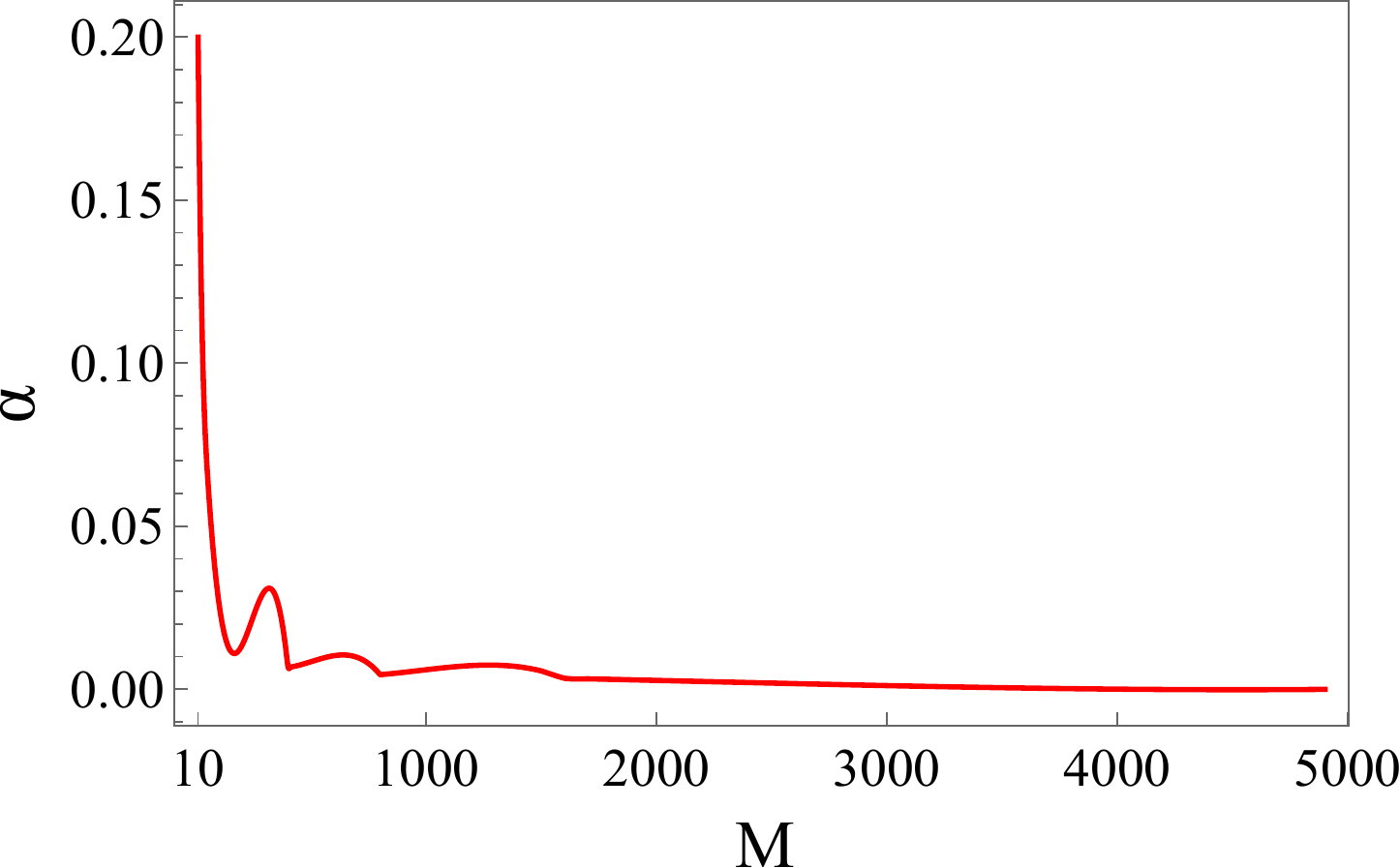}
    \caption{Inversion accuracy $\alpha$ as a function of M. Calculations are for a GNR of $L=100$ unit cells, all of which containing 3\% of impurities. }
    \label{error_M}
\end{figure}

Figure~\ref{error_M} confirms the earlier statement that $M=10^3$ is sufficient to achieve statistical significance and reduce fluctuations to acceptable levels. In fact, the inversion error practically vanishes once $M$ is of that order of magnitude. 

\subsection{5. Recursive Method} 
As shown in the main manuscript, the first order approximation of CA conductance 
induces a minimum in the misfit function located at $n^{*}=-B/A$. Both $A$ and $B$ result from integrations whose integrands involve the function $\beta(E)$ defined in the main text as the $n$-derivative of the CA conductance evaluated at $n=0$. While this is an excellent approximation for $n \leq 3\%$, it can be extended to higher concentrations (up to $7\%$) if we adopt the following recursive procedure. We must redefine $\beta(E)$ as the $n$-derivative of the CA conductance evaluated at an arbitrary concentration $n$ and follow the steps below:

\begin{enumerate}
    \item Evaluate average $\beta(E)$ between $n=0$ and an initial choice of $n$.
    \item Calculate $n^*$ using the expression given in the main manuscript with the average $\beta(E)$ function evaluated in step 1.
    \item Make $n=n^*$ and return to step 1 until $n^*$ converges. 
\end{enumerate}



\section{II. DFT calculations}

The DFT calculation was performed using a localized atomic orbital basis set (LCAO), implemented in the SIESTA package \cite{siesta}. All calculations were carried out using a double-$\zeta$ 
basis with polarization orbitals within the generalized gradient approximation (GGA-PBE) \cite{perdew1996generalized} for the exchange and correlation potential. A real space mesh cut-off of 300Ry, and a Monkhorst-Pack k-grid 80 points along the ribbon direction were chosen. The structures were fully relaxed down to a threshold force criteria of $0.01 eV/$\AA.
\begin{figure*}[t!]
    \twocolumngrid
    \includegraphics[width=1.75\columnwidth]{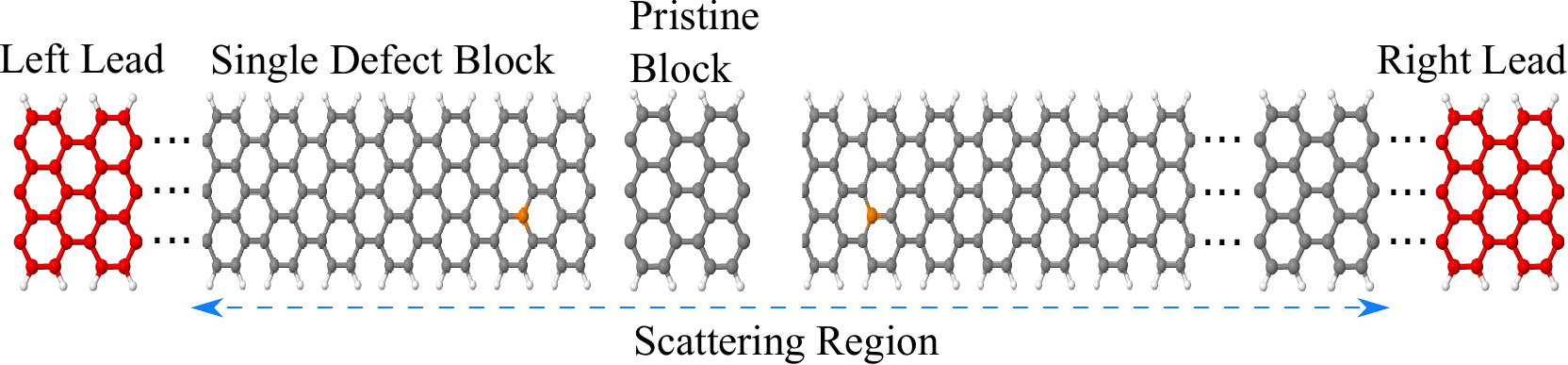}
\caption{ Illustrative diagram showing two different building blocks patched together at a random order to generate the finite-concentration device. The Hamiltonian of both blocks were obtained from DFT calculations.  }
    \label{plot:dft-sm}
\end{figure*}

Calculations for the disordered system were implemented in the following way: two initial DFT calculations were done, namely for a pristine GNR, and for a nanoribbon of length $L=7$ unit cells containing only a single nitrogen substitutional impurity. From the Hamiltonians of the pristine ribbon and the one containing a single nitrogen, we assemble the N-impurity device by randomly distributing these building blocks over a fixed length and defect concentration. This is illustrated in Fig.~\ref{plot:dft-sm} showing how the building blocks are patched together to form the device containing a finite concentration of impurities. Given the localized nature of the basis orbitals, the full Hamiltonian for the disordered problem is block tridiagonal. We can thus use a recursive method \cite{amorim2013,rocha2007,rocha2008,rocha2010} to find the Green function matrix elements required to compute the Landauer conductance. These were the steps followed to obtain the conductance spectra $\Gamma(E)$ of all DFT-based calculations. 

\subsection{1. DFT-based tight-binding Hamiltonian} 

There are numerous ways of extracting effective tight-binding (TB) Hamiltonians from DFT calculations, all of which reproduce the corresponding band structures very accurately \cite{gresch2017z2pack, Pizzi2020, nardelli2018paoflow}. Combined with the help of quantum transport scripts \cite{vanderbilt_2018}, the TB parameters can be exported to highly efficient codes \cite{groth2014kwant,Lima2018,joao2020kite} that can handle quantum transport calculations of very large systems. 
Despite the wide availability of such DFT-based TB Hamiltonians, for simplicity, the illustrative examples in the manuscript focus on single-band TB models. We stress, however, the following description is applicable to whichever TB-based model one chooses to adopt.
 
\begin{figure}
    \centering
    \includegraphics[width=\columnwidth,height =0.8\columnwidth]{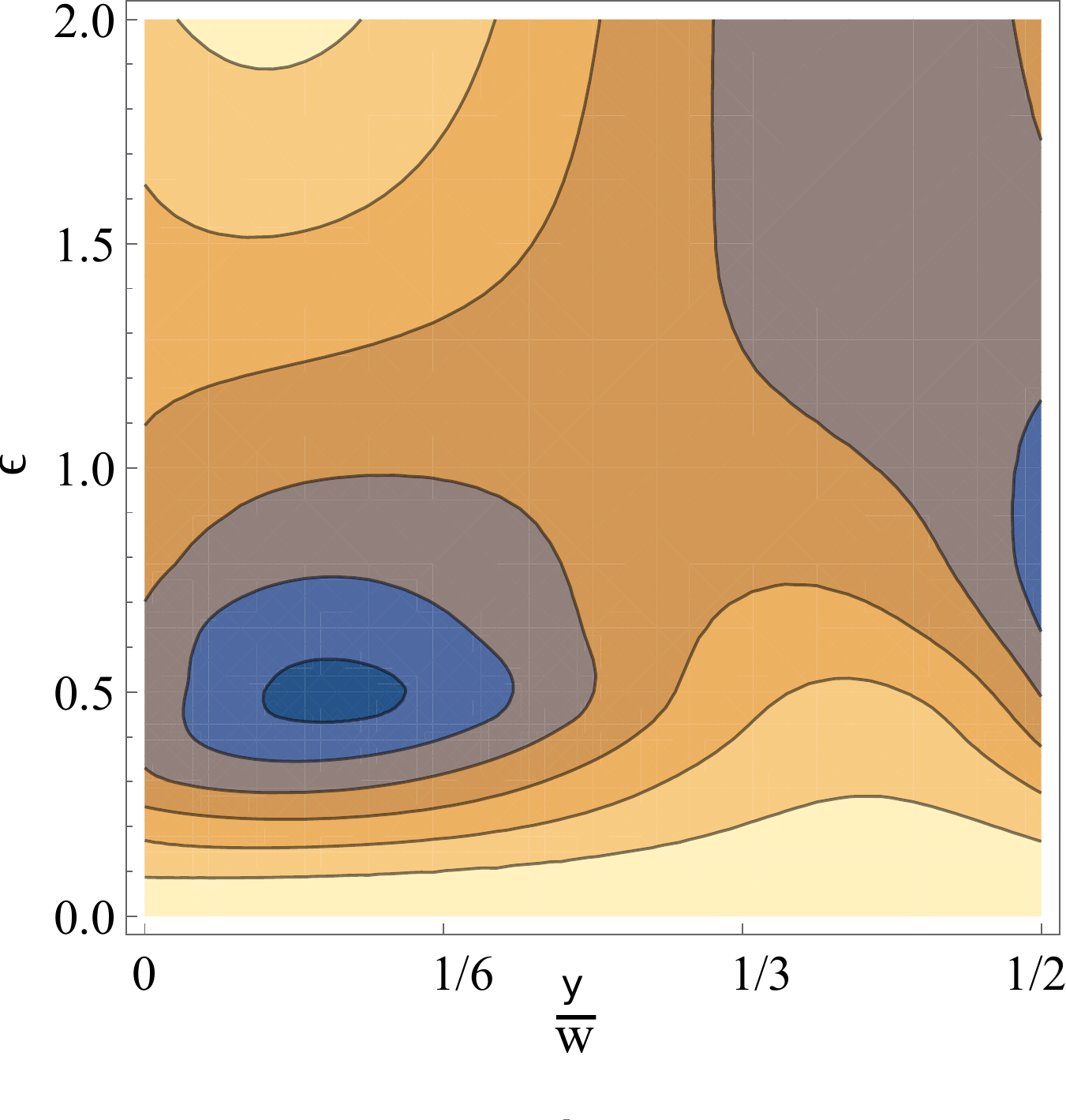}
    \caption{Contour plot of the misfit function $\chi$ to evaluate the impurity TB on-site energy $\epsilon$. The single impurity sample conductance contains information about the location as well as the scattering strength of the impurity. The variable $y/W$ indicates the impurity location, with $y/W = 0$  corresponding to the edge of the GNR and $y/W = 1/2$ marking its centre.
    }
    \label{TBscat}
\end{figure}

Instead of the usual band-structure fitting \cite{papaconstantopoulos1986handbook,slater1954simplified}, our adopted procedure consisted of comparing the conductance spectra calculated within DFT and tight-binding. For the sake of differentiation we refer to them as $\Gamma_N^{{\rm DFT}}$ and $\Gamma_N^{\rm TB}$, where the subscript simply indicates the number of impurities contained in the device. From the impurity-free results, {\it i.e.} $\Gamma_0^{\rm DFT}$ and $\Gamma_0^{\rm TB}$, it is straightforward to extract the hopping parameter $t$ that best captures the electronic structure of the pristine ribbon in question. The impurity on-site parameter $\epsilon$ is easily found from $\Gamma_1^{\rm DFT}$, which corresponds to the conductance in the presence of only a single impurity. It follows from a simple plot of the misfit function as a function of the impurity on-site potential, {\it i.e.} $\chi(\epsilon) = \int dE [\Gamma_1^{\rm DFT} - \Gamma_1^{\rm TB}(\epsilon)]^2$. In this case $\Gamma_1^{\rm TB}(\epsilon)$ is the tight-binding conductance of a ribbon with a single impurity with on-site potential $\epsilon$. The function $\chi(\epsilon)$ presents a very distinctive minimum, which we adopt as the value that best represents the scattering strength of the impurity. It is worth mentioning that it is not necessary to know the exact location of the impurity in the DFT calculation. In fact, this is automatically accounted for by evaluating the misfit function $\chi(\epsilon)$ for distinct impurity locations in the tight-binding part of the calculation. This is depicted in Fig.\ref{TBscat}. The absolute minimum found for $\chi(\epsilon)$ corresponds to the impurity location used in the DFT-based conductance which also defines the value of the on-site potential to be adopted in the subsequent CA calculation. 

\begin{figure}
    \centering
    \includegraphics[width=\columnwidth]{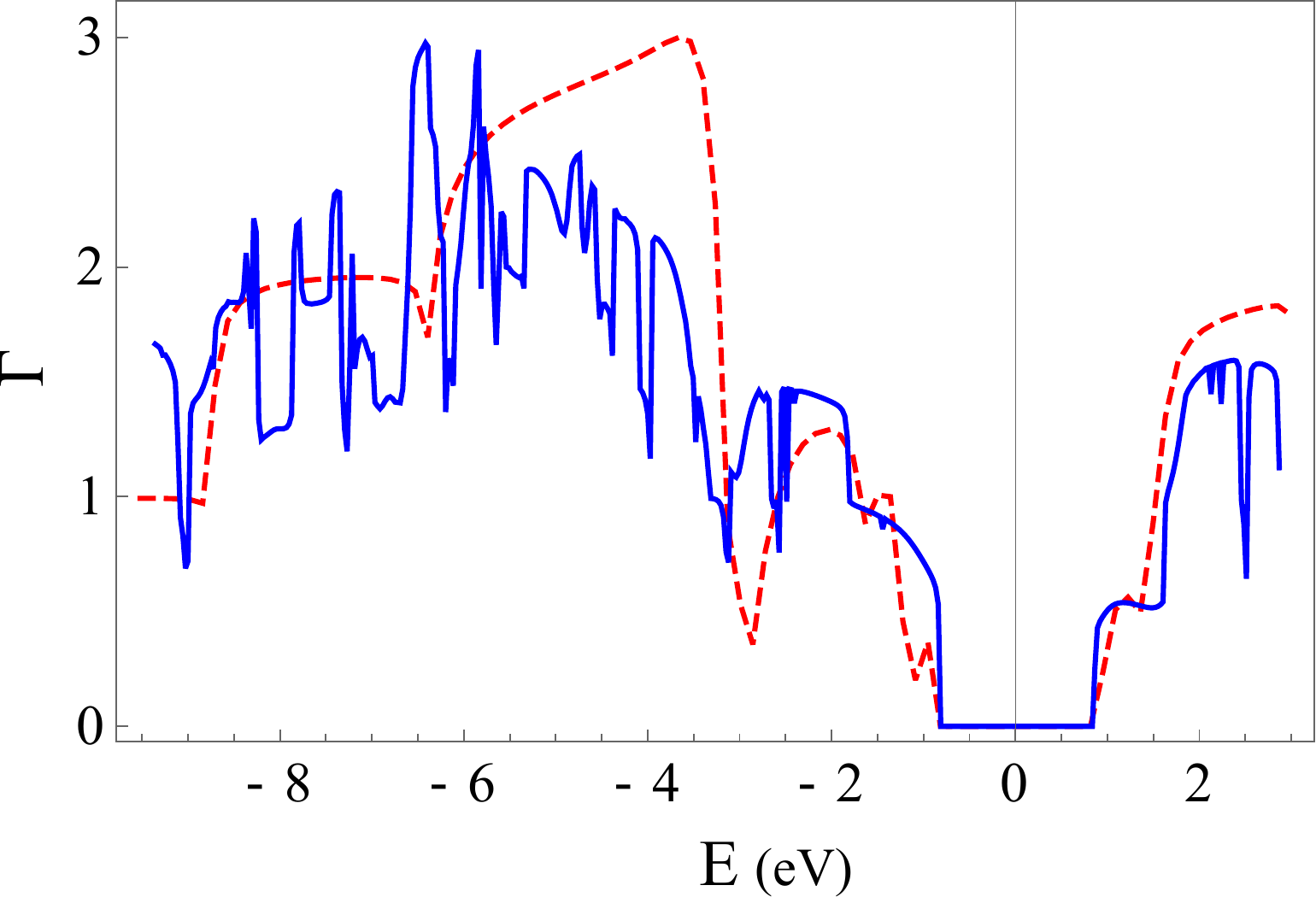}
    \caption{Conductance as a function of the energy (in eV). The (blue) solid line represents $\Gamma_1^{\rm DFT}$ and the (red) dashed line corresponds to $\Gamma_1^{\rm TB}$.}
    \label{TBscat2}
\end{figure}

Finally, it is also instructive to compare the conductance spectra $\Gamma_1^{\rm DFT}$ and $\Gamma_1^{\rm TB}$, the latter being calculated with the on-site potential $\epsilon$ obtained as described above. Bearing in mind that the misfit function was obtained by an integration in the $0-0.2 \, \rm{eV}$ energy range, the agreement between the two results shown in Fig.\ref{TBscat2} is yet another indication of the effectiveness of our inversion procedure since both curves were calculated separately and independently. 

\subsection{2. DFT-based misfit function}

\begin{figure}[h]
    \centering
    \includegraphics[width=\columnwidth]{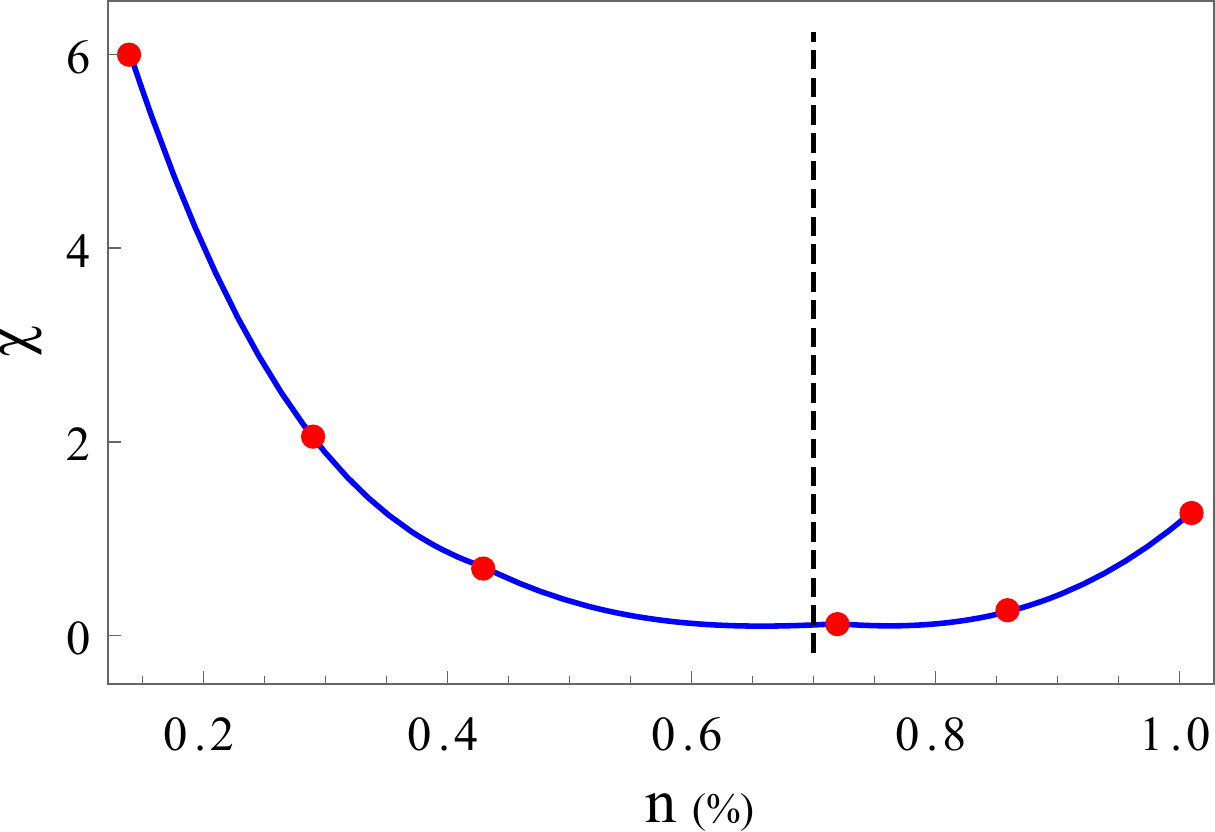}
    \caption{DFT based misfit function generated for an arbitrary configuration correctly indicates the minimum at 0.7\% shown as a black dashed line. }
    \label{dft_misfit}
\end{figure}

Although the approach described above is undoubtedly a good strategy to illustrate what can be achieved by carrying out the CA calculation within the TB model, it is by no means essential. To make this point evident, we have also carried out the CA part of the calculation entirely within DFT without availing of any TB parameterization. For each concentration value, M=50 different DFT disordered configurations were considered, each one of them generating conductance spectra with 500 energy values. The obtained misfit function $\chi(n)$ seen in Fig. \ref{dft_misfit} displays the same distinctive trend seen previously and with a minimum at exactly the same value ($n=0.7\%$) as the one depicted in Fig.4b of the main manuscript. This clearly indicates that the inversion method can decode the disorder signatures contained in the conductance signals regardless of the tool used to describe the underlying electronic structure. It is reassuring to find exactly the same answer using totally different tools.   

\bibliography{mybib,dftbib}